\documentclass[iop,apj]{emulateapj}
\usepackage{apjfonts}
\journalinfo{The Astrophysical Journal, 2013, in press}

\shorttitle{Alfv\'{e}n Wave Turbulence}
\shortauthors{Asgari-Targhi, van Ballegooijen, Cranmer, \& DeLuca}

\begin{document}

\title{The Spatial and Temporal Dependence of Coronal Heating by
Alfv\'{e}n Wave Turbulence}

\author{M. Asgari-Targhi,
A. A. van Ballegooijen,
S. R. Cranmer, and E. E. DeLuca}
\affil{Harvard-Smithsonian Center for Astrophysics,
60 Garden Street, Cambridge, MA 02138, USA} 

\begin{abstract}
The solar atmosphere may be heated by Alfv\'{e}n waves that propagate
up from the convection zone and dissipate their energy in the
chromosphere and corona. To further test this theory, we consider wave
heating in an active region observed on 2012 March 7. A potential
field model of the region is constructed, and 22 field lines
representing observed coronal loops are traced through the model.
Using a three-dimensional (3D) reduced magneto-hydrodynamics (MHD)
code, we simulate the dynamics of Alfv\'{e}n waves in and near
the observed loops. The results for different loops are combined into
a single formula describing the average heating rate $Q$ as function
of position within the observed active region. We suggest this
expression may be approximately valid also for other active regions,
and therefore may be used to construct 3D, time-dependent models of
the coronal plasma. Such models are needed to understand the role
of thermal non-equilibrium in the structuring and dynamics of the
Sun's corona.
\end{abstract}


\section{Introduction}
\label{sect:Intro}
The solar corona is two orders of magnitude hotter than the underlying
photosphere. This has been puzzling scientists over many decades as
they try to identify the mechanism(s) behind these extreme conditions.
Many different ideas on coronal heating have been proposed \citep[see
reviews by][]{Aschwanden2005, Klimchuk2006}, but solar observations
have not yet advanced to the point where they allow us to clearly
identify the physical mechanisms responsible for coronal heating. 
One plausible idea is that the corona is heated by dissipation of
Alfv\'{e}n waves \citep[][]{Hollweg1981, Hollweg1986, Heyvaerts1983, Kudoh1999,
Moriyasu2004, Matsumoto2010, Antolin2010} or by turbulence
\citep[][]{Rappazzo2008, Dmitruk1997, Cranmer2007}. Alfv\'{e}n waves have
indeed been observed at various heights in the solar atmosphere
\citep[][]{Ulrich1996, Fujimura2009, DePontieu2007a, Tomczyk2009,
McIntosh2011}, but the role of such waves in coronal heating has not
yet been clearly demonstrated.

Recently, the present authors developed a 3D MHD model describing the
propagation and dissipation of Alfv\'{e}n waves in a coronal loop,
including the lower atmospheres at the two ends of the loop
\citep[][hereafter paper I]{vanB2011}. According to this model, the
waves are generated by the interactions of granule-scale convective
flows within kilogauss flux tubes in the photosphere. The flows
produce non-axisymmetric kink-type waves on a transverse length scale
less than 100 km in the photosphere, below the resolution of
present-day telescopes. The waves propagate upward along the tubes and
strongly reflect in the chromosphere and transition region (TR),
producing counter-propagating waves that are subject to strong
nonlinear wave--wave interactions 
\citep[e.g.,][]{Shebalin1983, Higdon1984, Oughton1995, Goldreich1995,
Goldreich1997, Bhattacharjee2001, Cho2002, Oughton2004}.
These nonlinear interactions result in turbulent transfer of wave energy to small
spatial scales and heating of the chromospheric plasma. A fraction of
the wave energy is transmitted through the TR into the corona, and
produces turbulence and heating there. It was found that this model
can quantitatively explain both the chromospheric and coronal heating
rates observed in active regions, and is consistent with observational
constraints on the degree of braiding of the coronal loops.

A key feature of the Alfv\'{e}n wave turbulence model is that it
naturally predicts the spatial variation of heating rate $Q(s)$ with
position $s$ along a coronal loop, so the heating profile is not
arbitrarily prescribed. The thermal stability of a coronal loop
depends strongly on the heating profile \citep[e.g.][]{Antiochos1991,
Muller2003, Testa2005, Karpen2006, Mok2008, Klimchuk2010}. If most of
the heat is deposited near the loop footpoints, much of the energy is
conducted downward into the upper chromosphere, producing strong
chromospheric evaporation and large coronal densities. The radiative
losses at the loop top may then become too large to be balanced by local
heating. This results in the formation of cool condensations that move
along the loop under the influence of gravity and pressure gradients.
Modeling predicts that the loop undergoes periodic convulsions as it
searches for a nonexistent equilibrium state. \citet[]{Klimchuk2010}
examine the possibility that the observed warm (1--1.5 MK) loops in
active regions can be explained by such thermal nonequilibrium.

\citet[][hereafter paper II]{Asgari2012} applied the Alfv\'{e}n wave
turbulence model to an active region observed with the Solar Dynamics
Observatory (SDO). The dynamics of Alfv\'{e}n waves in selected coronal
loops was computed by the model of paper I, based on magnetic
field strengths taken from a 3D magnetic model of the active region.
Wave heating rates were derived from these simulations, and average
coronal temperatures and densities were computed self-consistently
with the modeled heating rates. It was found that the loops in the
core of the active region were thermally stable with peak temperatures
of 2--3 MK. The expected temperature fluctuations of such loops are
quite small, $\Delta T \sim 0.1$ MK, implying that the loops never
cool off into the 1--1.5 MK range. However, other loops at the
periphery of the active region were predicted to be thermally
unstable, and presumably exist in a state of thermal nonequilibrium
with large temperature and density fluctuations. The thermal evolution
of such unstable loops was not addressed in paper II.

To further test the Alfv\'{e}n wave turbulence model, it is necessary
to construct detailed numerical models of wave-heated active regions
and then compare such models with solar observations. Such models must
describe not only the magnetic field but also the temperature and
density of the plasma. Both thermally stable and unstable coronal
loops must be considered. To follow the evolution of the unstable
loops, the time evolution of temperature and density must be simulated
with high spatial resolution as neighboring loops evolve almost
independently. Such simulations will be very challenging and are
beyond the scope of the present paper. However, the simulations will
require detailed knowledge of how the local wave heating rate depends
on coronal field strength, loop length, and coronal density. It is
also important to know how the heating rate varies with position along
a coronal loop because this is an important factor in determining
whether a loop is stable or unstable.

The main objective of the present paper is to derive an expression for
the wave heating rate that can be used in future 3D MHD simulations of
active region structure and evolution. A second objective is to study
the internal motions of a wave-heated loop and discuss the effects of such
motions on plasma temperature. These
results are important for understanding the observable consequences of
the Alfv\'{e}n wave turbulence model.

Our previous work indicates that the wave heating rate $Q$ depends
strongly on magnetic structure, and to lesser extent on coronal
density (see papers I and II). Therefore, it is important that the
Alfv\'{e}n wave simulations are done using a realistic model for the
background magnetic field. In the present paper we use data from the
Helioseismic and Magnetic Imager (HMI) on SDO to construct a 3D
potential field model of an observed active region, and we trace
22 field lines through this model. For each field line we construct a
detailed numerical model of the wave dynamics, which yields the
time-averaged heating rate $Q(s)$ as function of position along that
field line. Finally, we combine the results from different field lines
into a single expression for the heating rate and its dependence on
magnetic field strength, density and the loop length.

The paper is organized as follows. Section 2 presents the SDO
observations of an active region, and the construction of the 3D
magnetic model for this region. Section 3 describes the Alfv\'{e}n wave
turbulence modeling for individual flux tubes. Section 4 describes the
simulation results. Section 5 presents the combined formula for the
wave heating rate. Section 6 discusses the internal wave dynamics. The modeling results are further discussed in Section 7.

\section{Observations and Potential Field Modeling}

\begin{figure}
\epsscale{1.01}
\plotone{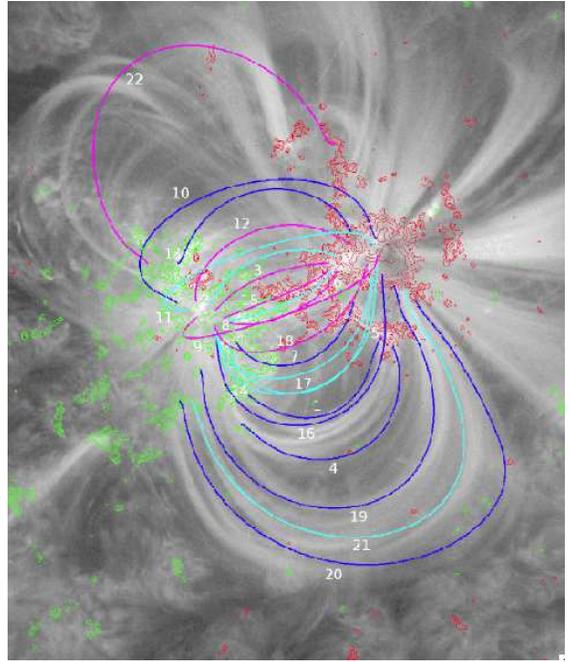}
\caption{SDO observation of active region NOAA 11428 on 2012 March 7
at 18:30 UT with field of view of 276 $\times$ 315 Mm.
The image was taken with AIA in the 171~{\AA} band and uses
logarithmic scaling. The red and green contours show the magnetic flux
distribution in the photosphere, as derived from HMI magnetograms
assuming the field is radial (red is positive, green is negative,
contours at 79, 159, 318, 636 and 1271 G). The blue and magenta curves
are magnetic field lines traced through the PFSS model.}
\label{fig1}
\end{figure}

The active region used in the present study is NOAA 11428, which
crossed the central meridian on 2012 March 7 and has a relatively
simple, bipolar magnetic structure. We use SDO data taken around
18:30 UT on that day. Figure~\ref{fig1} shows the magnetic flux
distribution in the photosphere (red and green contours) superposed
on the AIA 171 {\AA} image from 18:30:01 UT. The flux distribution
is derived from HMI line-of-sight magnetograms taken over a 15 minute
period. The method for extracting the flux distribution from a
magnetogram is described by \citet[]{Su2009a}. Sunspots are 
present in both the leader and follower parts
of the region. The AIA image shows a set of coronal loops and fans
emanating from the two spots and surrounding fluxes. 

This particular active region does not exhibit much ``sigmoidal twist'',
therefore it is modeled based on  Potential Field Source Surface (PFSS) model. 
We construct a PFSS model of the observed region, 
based on the flux distribution derived from the HMI
magnetogram and the SOLIS synoptic map for Carrington Rotation 2121.
The PFSS model was computed using the Coronal Modeling System software
\citep[][]{Su2011}. The model includes both a high resolution grid
covering the target active region (grid spacing $10^{-3}$
$R_\odot$) and a low resolution grid covering the entire Sun
($1^\circ$ spacing). The source surface is assumed to be located at
radial distance $r_s = 2.05 \, R_\odot$ from Sun center. We trace 22
randomly selected field lines through the PFSS model.
Figure \ref{fig1} shows the selected field lines superposed on the AIA
171 {\AA} image. For each field line we determine the magnetic field
strength $B(s)$ and the height $z(s)$ above the photosphere as
functions of position $s$ along the field line, which are needed for
the wave modeling described in the next section.

\section{Alfv\'{e}n Wave Turbulence Model}
\label{sect:Model}

The Alfv\'{e}n waves are simulated using the 3D MHD model developed in
papers I and II. In this model a thin flux tube surrounding a traced
magnetic field line is considered. The tube is assumed to have a
circular cross-section with radius $R(s)$ that varies with position
$s$ along the tube, which extends from the photosphere at one end of
the coronal loop to the photosphere at the other end. Even though our
PFSS model uses high spatial resolution, the field strengths
in the lower atmosphere predicted from the magnetogram 
are not as high as those known to exist on the
Sun \citep[][]{Stenflo1973}. Therefore, we artificially increase the field strengths at the
ends of the flux tube to produce kilogauss field strengths in the
photosphere: $B_{\rm Phot} =1400$ G. This is done on each computed field
line, but only at heights below the TR. The corrected field strength
is 
\begin{equation}
  B(s)  = \sqrt {B_{\rm Phot} ^2 \tilde{f}(s) + B_0^2 (s) \left[1 - \tilde{f}(s)
    \right] } ,
 \label{eq:bs}
\end{equation}
where  $\tilde{f}(s)$ is a weighting factor derived from the gas pressure:
\begin{equation}
  \tilde{f}(s)  = \frac{p(s) - p_{\rm TR}}{p_{\rm Phot} - p_{\rm TR}} ,
 \label{eq:weit}
\end{equation}
and $B_0(s)$ is the field strength as derived from the PFSS model. The
height dependent gas pressure $p(s)$ decreases monotonically from its
photospheric value ($p_{\rm Phot}$) to its value at the TR ($p_{\rm TR}$).
The background field strength $B(s)$ and plasma density
$\rho (s)$ are assumed to be constant over the tube's cross-section,
so the Alfv\'{e}n speed $v_A (s)$ ($\equiv B / \sqrt{4 \pi \rho}$) is
also constant over the cross-section. The plasma temperature $T(s)$ is
fixed in time, and mass flows along the flux tube are neglected.
Therefore, in its present form the model is not suitable for
simulating thermally unstable loops, which have time-variable temperature,
density and Alfv\'{e}n speed.

At the two ends of the flux tube the field lines are subjected to
random footpoint motions that simulate the effects of convective flows
on kilogauss flux elements in the photosphere. The imposed footpoint
motions are assumed to have a root-mean-square (rms) velocity of
1.5 $\rm km ~ s^{-1}$ and correlation time of 60 s, similar to the
observed random motions of G-band bright points in the photosphere
\citep[e.g.,][] {Chitta2012}. The footpoint motions produce Alfv\'{e}n
waves that travel upward inside the flux tube and reflect at various
heights due to spatial variations in Alfv\'{e}n speed $v_A (s)$. Such
reflections result in counter-propagating Alfv\'{e}n waves that
interact nonlinearly with each other and produce wave turbulence \citep[][]
{Kraichnan1965,Shebalin1983, Oughton1995, Cho2002, Cranmer2010}.

The wave dynamics are simulated numerically using the ``reduced MHD''
approximation \citep[e.g.,][paper I]{Strauss1976},
which means that slow- and fast MHD modes are filtered out and only
the Alfv\'{e}n mode is retained. Furthermore, only the magnetic- and
velocity fluctuations of the waves are simulated, not their effects
on temperature and density. The magnetic fluctuations
${\bf B}_1$ are assumed to be small compared to the background field
${\bf B}$, and can be approximated as ${\bf B}_1 = \nabla_\perp h
\times {\bf B}$, where $h({\bf r},t)$ is the magnetic flux function
and $t$ is the time. Similarly, the velocity fluctuations are
approximated as ${\bf v}_1 = \nabla_\perp f \times \hat{\bf B}$,
where $f({\bf r},t)$ is the velocity stream function and
$\hat {\bf B} (x,y,s)$ is the unit vector along the background
field. Flows along the background field are neglected. In the present
version of the reduced MHD model the functions $f({\bf r},t)$ and
$h({\bf r},t)$ satisfy the following coupled equations:
\begin{equation}
\frac{\partial \omega} {\partial t} + \hat{\bf B} \cdot
( \nabla_\perp \omega \times \nabla_\perp f ) = v_A^2 \left[
\hat{\bf B} \cdot \nabla \alpha + \hat{\bf B} \cdot
( \nabla_\perp \alpha \times \nabla_\perp h ) \right] + D_v , 
\label{eq:dodt}
\end{equation}
\begin{equation}
\frac{\partial h} {\partial t} = \hat{\bf B} \cdot \nabla f
+ \frac{f} {H_B} + \hat{\bf B} \cdot ( \nabla_\perp f \times
\nabla_\perp h ) + D_m ,
 \label{eq:dhdt}
\end{equation}
where $\omega \equiv - \nabla_\perp^2 f$ is the parallel component
of vorticity, $\alpha \equiv -\nabla_\perp^2 h$ is the magnetic
torsion parameter, and $H_B (s) \equiv B/(dB/ds)$ is a magnetic
scale length. The terms $D_v$ and $D_m$ describe the effects of
viscosity and resistivity on the high wavenumber modes. Derivations
of the above equations and descriptions of their numerical
implementation are given in paper~I.

The spatial dependence of the waves on the transverse coordinates $x$ and $y$ is
described in terms of the eigenfunction of the $\nabla_\perp^2$
operator on the domain $x^2 + y^2 < R^2$. These eigenfunctions have
perpendicular wavenumbers $k_\perp = a_k /R(s)$, where the $a_k$ are
given by the zeros of Bessel functions, and $k$ is the mode index
($k = 1, \cdots, N$). In the present work we use $a_{\rm max} = 30$,
which corresponds to a spatial resolution of about one-tenth of the
tube radius and requires $N = 209$ modes. Also, the viscous and
resistive damping rates are assumed to be equal, and vary with the
fourth power of the perpendicular wavenumber (``hyperdiffusion''): 
\begin{equation}
\nu_k (s,t) = \nu_{\rm max} (s,t) \left( \frac{a_k} {a_{\rm max}}
\right)^4 . \label{eq:nuk}
\end{equation}
The maximum damping rate is given by $\nu_{\rm max} = 70 ~
\overline{v_{\rm rms}}(s,t) /R(s)$, where $v_{\rm rms} (s,t)$ is the
rms velocity of the waves, and the bar represent a running time
average over a time interval of 300~s.

In the TRs at the two ends of the coronal loop the temperature
and density vary rapidly with position, causing strong variation in
Alfv\'{e}n speed and therefore strong wave reflection. To avoid having
to use very small time steps, we treat the TRs as discontinuities
where the waves can reflect. In the corona the temperature is assumed
to be a function of height only:
\begin{equation}
T(s) = T_{\rm max} \left[ 0.01 + 0.99 \left( \frac{z(s) -
z_{\rm TR}} {z_{\rm max} - z_{\rm TR}} \right) \right]^{2/7} ,
\label{eq:T}
\end{equation}
where $z(s)$ is the height as derived from the PFSS model,
$z_{\rm max}$ is the maximum height, and $z_{\rm TR}$ is the height of
the two TRs. The peak temperature $T_{\rm max}$ is estimated from the
coronal loop length $L$ and coronal pressure $p$, using the RTV
scaling law \citep[][]{Rosner1978}. The coronal pressure $p$ is the
only free parameter of the model. The density $\rho (s)$ is computed
from the temperature by solving the equation of hydrostatic
equilibrium along the field line.

The way we have set up the model is such that the temperature
and density are fixed in time, so the radiative and conductive losses
are also fixed. However, these radiative and conductive loss rates are
not necessarily equal to the wave heating rate $Q(s)$, which is
determined by averaging $Q(s,t)$ over the duration of the
simulation. Therefore, the loops simulated here are generally not in
thermal equilibrium. When the heating rate is higher (lower) than the
radiative and conductive losses, the coronal temperature is expected
to increase (decrease) with time, but this variation with time is not
taken into account in our simulations.  
  
After Equations (\ref{eq:dodt}) and (\ref{eq:dhdt}) have been solved,
the volumetric heating rate can be computed as a sum of viscous- and
resistive contributions: $Q = Q_{\rm kin} + Q_{\rm mag}$. The averages
of these quantities over the loop cross-section can be written as sums
over perpendicular modes:
\begin{eqnarray}
Q_{\rm kin} (s,t) & = & \frac{\rho}{R^2}
\sum_{k=1}^N \nu_k a_k^2 f_k^2 , \label{eq:Qkin1} \\
Q_{\rm mag} (s,t) & = & \frac{B^2}{4 \pi R^2}
\sum_{k=1}^N \nu_k a_k^2 h_k^2 , \label{eq:Qmag1}
\end{eqnarray}
where $f_k (s,t)$ and $h_k (s,t)$ are mode amplitudes for the velocity
stream function and magnetic flux function, respectively (see Appendix
C in paper~I). In this paper we also compute the local heating rate
$Q(x,y,s,t)$ at any point within the loop. This quantity is the sum
of the following non-negative expressions:
\begin{equation}
Q_{\rm kin} (\xi,\varphi,s,t)  =  \frac{\rho}{R^2} \nu_{\rm max}
\left[ \sum_{k=1}^N \left( \frac{a_k}{a_{\rm max}} \right)^2 a_k f_k
F_k (\xi,\varphi) \right]^2 , \label{eq:Qkin2}
\end{equation}
\begin{equation}
\!\!\! Q_{\rm mag} (\xi,\varphi,s,t) = \frac{B^2}{4 \pi R^2} \nu_{\rm max}
\left[ \sum_{k=1}^N \left( \frac{a_k}{a_{\rm max}} \right)^2 a_k h_k 
F_k (\xi,\varphi) \right]^2 , \label{eq:Qmag2}
\end{equation}
where $\xi \equiv r/R$ is a dimensionless radial coordinate, $\varphi$
is the azimuth angle, and $F_k (\xi,\varphi)$ is a set of orthogonal
basis functions (see Appendix B in paper~I). Integrating these
expressions over the cross-section and using the orthogonality of the
eigenmodes, Equations (\ref{eq:Qkin1}) and (\ref{eq:Qmag1}) are
recovered.

\section{Simulation Results}

The above turbulence modeling was applied to 22 field lines traced
through the PFSS model for NOAA region 11428, where each field line
represents a particular coronal loop. The PFSS model predicts that
most loops expand significantly with height; the loop expansion factor
is defined by $\Gamma \equiv B_{\rm TR} / B_{\rm min}$, where
$B_{\rm TR}$ is the magnetic field strength at the TR (average of the
two TRs) and $B_{\rm min}$ is the minimum field strength along the
loop. For each loop we construct several models for the background
atmosphere with different values of the coronal pressure $p$, leading
to a total of 77 different models shown in Table 1. The models use a
broad range of plasma pressures because the results will be used to 
construct a heating rate formula (see section 5) to be used in future 
dynamical loop modeling. We anticipate that for thermally unstable
loops the temperature and density will vary strongly with time, and 
our formula should capture the effects of density on the wave heating
rate. For thermally stable loops the highest value of pressure was
chosen to lie above the value for thermal equilibrium 
(i.e., wave heating balanced by radiative and conductive losses).

\begin{deluxetable*}{crccccccccc}
\tablenum{1}
\tablewidth{0pt}
\tablecaption{Parameters of the Loops \label{table3}}
\tablehead{
\colhead{Model} & \colhead{$L$} & \colhead{$z_{\rm TR}$} & \colhead{$\Gamma$} & \colhead{$B_{\rm min}$}
&\colhead{$B_{\rm TR}$} &\colhead{$p$} & \colhead{$\tau_{AC}$} &\colhead{$m$} &\colhead{$Q_{\rm TR}$} &
 \colhead{$\eta_{Q}$} \\
\colhead{} & \colhead{[Mm]} & \colhead{[km]} & \colhead{} & \colhead{[G]}
& \colhead{[G]} & \colhead{[$\rm dyne/cm^{2}$]} & \colhead{s} &
\colhead{} & \colhead{[$\rm erg/cm^{3}/s$]}
&\colhead{} }
\startdata
f1r1   & 55.7 & 3229 & 2.9   & 37.0   & 106.3  & 0.25 &15.78 &0.533   &$4.094\times10^{-3}$& 1.75  \\
f1r2   & 59.8 & 2623 & 3.5   & 37.0   & 128.6  & 1.00 &25.70 &0.556   &$5.580\times10^{-3}$& 2.00 \\
f1r3   & 63.7 & 2013 & 4.4   & 37.0   & 162.4  & 4.00 &41.42 &0.507   &$6.787\times10^{-3}$&2.12\\
f1r4   & 64.2 & 1921 & 4.6   & 37.0   & 169.8  & 5.00 & 44.77&0.534   &$7.698\times10^{-3}$&2.26\\
\hline
f2r1   & 59.6 & 3232 & 5.6   &  86.8  & 487.8  &0.25  &5.85   &0.672   &$5.778\times10^{-3}$ & 3.19 \\
f2r2   & 61.5 & 2624 & 6.4   &  86.8  & 551.5  &1.00  &9.60   &0.663   &$7.214\times10^{-3}$ &3.41\\
f2r3   & 63.2 & 2016 & 7.2   &  86.8  & 621.9  &4.00  &15.56 &0.642   &$9.884\times10^{-2}$&3.54\\
f2r4   & 63.7 & 1845 & 7.4   &  86.8  & 641.8  &6.00  &17.90 &0.632   &$1.147\times10^{-2}$&3.54\\
\hline
f3r1   & 51.6 & 3229 & 4.8   & 106.9 & 507.8  &0.25  &4.35  &0.661    &$5.473\times10^{-3}$ &2.80\\
f3r2   & 53.2 & 2625 & 5.3   & 106.9 & 569.4  &1.00  &7.14  &0.651    &$7.694\times10^{-3}$&2.97\\
f3r3   & 54.8 & 2021 & 6.0   & 106.9 & 641.2  &4.00  &11.55&0.630    &$9.701\times10^{-2}$&3.09\\
f3r4   & 55.3 & 1840 & 6.2   & 106.9 & 666.7  &6.00  &13.32&0.601    &$1.122\times10^{-2}$ &3.01\\
\hline
f4r1   & 79.8 & 3236 & 3.8   & 18.1   & 68.1  & 0.25  &40.32 &0.694   &$1.700\times10^{-3}$&2.50\\
f4r2   & 83.7 & 2633 & 4.2   & 18.1   & 76.0  & 1.00  &64.37 &0.555   &$1.837\times10^{-3}$&2.21\\
f4r3   & 86.8 & 2024 & 4.4   & 18.1   & 80.1  & 4.00  &103.57&0.632   &$ 2.224\times10^{-3}$&2.55\\
\hline
f5r1   & 83.3 & 3236 & 5.1   & 43.4  & 221.7 & 0.25  &14.41 &0.664    &$3.726\times10^{-3} $&2.95\\
f5r2   & 86.7 & 2628 & 6.4   & 43.4  & 276.1 & 1.00  &24.13 &0.584    &$ 4.963\times10^{-3} $&2.94\\
f5r3   & 90.1 & 2023 & 8.1   & 43.4  & 350.9 & 4.00  &39.54 &0.535    &$6.521\times10^{-3} $&3.06\\
\hline
f6r1   & 49.2 & 3236 & 4.0   &116.0   &471   &0.25   &3.97  &0.617    &$4.924\times10^{-3}$&2.37\\
f6r2   & 50.9 & 2624 & 4.7   &116.0   &544   &1.00   &6.60  &0.635    &$6.896\times10^{-3}$&2.67\\
f6r3   & 52.6 & 2021 & 5.5   &116.0   &639.8 &4.00  &10.65  &0.657    &$9.308\times10^{-3}$&3.07\\
\hline
f7r1   & 57.0 &3238 & 3.14   &74.2    &233.1 &0.25  &7.24 &0.601    &$3.592\times10^{-3}$&2.00\\
f7r2   & 58.9 &2633 & 3.46   &74.2    &256.8 &1.00  &12.00&0.645    &$ 5.105\times10^{-3}$&2.23\\
f7r3   & 60.7 &2023 & 3.83   &74.2    &283.9 &4.00  &19.74&0.697    &$ 6.584\times10^{-3}$&2.55\\
\hline
f8r1   & 47.0 &3231 & 4.0    &121.9   &481.9 &0.25   &3.68 &0.605    &$5.445\times10^{-3}$&2.30\\
f8r2   & 48.6 &2628 & 4.4    &121.9   &541.5 &1.00   &6.04 &0.620    &$7.579\times10^{-3}$&2.52\\
f8r3   & 50.2 &2012 & 5.1    &121.9   &619.6 &4.00   &9.89 &0.647    &$1.001\times10^{-2}$&2.86\\
\hline
f9r1   & 175.9&3230 & 61.7  & 12.1   & 746.1  & 0.25&52.19 & 0.696   $^\ast$ &$1.913\times 10^{-3}$ & 17.59\\
f9r2   & 176.5&2926 & 64.6  & 12.1   & 780.9  & 0.50&71.07 &0.787   $^\ast$ &$2.204\times 10^{-3}$ & 26.52\\
f9r3   & 177.2&2622 & 67.6  & 12.1   & 817.5  & 1.00&95.32 &0.814   $^\ast$ &$2.842\times 10^{-3}$ & 30.87\\
f9r4   & 177.8&2319 & 71.0  & 12.1   & 857.6  & 2.00&126.23 &0.822   $^\ast$ &$3.724\times 10^{-3}$ & 33.15\\
f9r5   & 178.4&2017 & 74.4  & 12.1   & 900.4  & 4.00&165.47 &0.820   $^\ast$ &$4.155\times 10^{-3}$ & 34.24\\
\hline
f10r1 & 211.3 & 3228 & 83.3 & 6.75   & 562.3  & 0.25 &103.56 &0.817   $^\ast$  &$1.7001\times 10^{-3}$ &37.04\\
f10r2 & 212.1 & 2920 & 89.2 & 6.75   & 601.5  & 0.50 &141.38 &0.885   $^\ast$  &$2.365\times 10^{-3}$ & 53.28\\
f10r3 & 212.8 & 2627 & 95.3 & 6.75   & 642.9  & 1.00 &187.87&0.932   $^\ast$  &$3.149\times 10^{-3}$ & 69.95\\
f10r4 & 213.6 & 2319 & 102.3& 6.75   &690.2  & 2.00 &249.58 &0.907   $^\ast$ &$3.476\times 10^{-3}$ & 66.47\\
f10r5 & 214.3 & 2020 & 109.4& 6.75   &738.3  & 4.00 &325.91 &0.874   $^\ast$ &$3.506\times 10^{-3}$ & 60.68\\
\hline
f11r1 & 186.5 & 3232 & 68.2 & 10.0   & 682.2  & 0.25 &65.16 &0.764   $^\ast$ &$1.573\times 10^{-3}$ & 25.17 \\
f11r2 & 187.1 & 2926 & 71.3 & 10.0   & 713.3  & 0.50 &88.63 &0.762   $^\ast$ &$2.113\times 10^{-3}$ & 25.87 \\
f11r3 & 187.8 & 2623 & 74.5 & 10.0   & 745.4  & 1.00 &118.79&0.816   $^\ast$ &$2.689\times 10^{-3}$ &33.64 \\
f11r4 & 188.4 & 2320 & 77.9 & 10.0   & 779.3  & 2.00 &157.25&0.856   $^\ast$ &$3.643\times 10^{-3}$ &41.65 \\
f11r5 & 189.0 & 2017 & 81.0 & 10.0   & 811.2  & 4.00 &206.07 &0.831   $^\ast$ &$3.685\times 10^{-3}$ &38.50 \\
\hline
f12r1 & 84.2  & 3233 & 10.0 & 51.5   & 514.9  & 0.25 &11.41 &0.743   &$3.678\times10^{-3}$&5.53\\
f12r2 & 85.9  & 2623 & 11.4 & 51.5   & 586.1  & 1.00 &19.01 &0.602   &$5.294\times10^{-3}$&4.32\\
f12r3 & 87.6  & 2015 & 13.0 & 51.5   & 667.8  & 4.00 &31.07 &0.561   &$6.518\times10^{-3}$&4.21\\
\hline
f13r1 & 91.6  & 3226 & 6.8  & 33.8   & 230.6  & 0.25 &19.33 &0.629   &$2.893\times10^{-3}$&3.35\\
f13r2 & 92.7  & 2930 & 7.3  & 33.8   & 245.9  & 0.5  &24.76 &0.612   &$3.267\times10^{-3}$&3.37\\
f13r3 & 93.8  & 2626 & 7.7  & 33.8   & 261.6  & 1.00 &31.82 &0.627   &$3.946\times10^{-3}$&3.60\\
f13r4 & 95.9  & 2015 & 8.7  & 33.8   & 293.7  & 4.00 &52.33 &0.601   &$5.033\times10^{-3}$&3.66\\
\hline
f14r1 & 115.2 & 3230 & 19.4 & 29.6   & 572.8  & 0.25 &21.00&0.623   &$3.099\times10^{-3}$&6.35\\
f14r2 & 116.9 & 2623 & 22.3 & 29.6   & 660.3  & 1.00 &35.99&0.705   &$4.363\times10^{-3}$&8.95\\
f14r3 & 118.6 & 2018 & 25.9 & 29.6   & 764.1  & 4.00 &59.85&0.667   &$5.330\times10^{-3}$&8.75\\
\hline
f15r1 & 47.3  & 3225 & 4.2  & 121.2 & 508.6   & 0.25 &3.66 &0.602   &$5.606\times10^{-3}$ &2.37\\
f15r2 & 48.9  & 2617 & 4.7  & 121.2 & 574.9   & 1.00 &6.05 &0.642   &$7.917\times10^{-3}$ &2.72\\
f15r3 & 50.4  & 2011 & 5.4  & 121.2 & 653.3   & 4.00 &9.86 &0.605   &$1.077\times10^{-2}$ &2.77\\
f15r4 & 50.8  & 1847 & 5.6  & 121.2 & 674.5   & 6.00 &11.23&0.607   &$1.116\times10^{-2}$ &2.83\\
\hline
f16r1 & 127.6 & 3233 & 17.1 & 22.1  & 378.8   & 0.25 &30.92&0.659   &$2.525\times10^{-3}$ &6.50\\
f16r2 & 130.2 & 2627 & 21.0 & 22.1  & 463.7   & 1.00 &52.64&0.738   &$3.681\times10^{-3}$ &9.45\\
f16r3 & 132.6 & 2019 & 25.5 & 22.1  & 564.6   & 4.00 &87.51 &0.754  &$4.939\times10^{-3}$ &11.52\\
\hline
f17r1 & 107.8 & 3230 & 18.4 & 34.5  & 635.1   & 0.25 &17.16 &0.655   &$2.778\times10^{-3}$ &6.75\\
f17r2 & 109.4 & 2623 & 21.0 & 34.5  & 723.3   & 1.00 &29.39 &0.660   &$4.115\times10^{-3}$ &7.44\\
f17r3 & 111.0 & 2017 & 24.0 & 34.5  & 827.3   & 4.00 &48.93 &0.699   &$5.799\times10^{-3}$ &9.21
\enddata               
\end{deluxetable*}

\begin{deluxetable*}{crccccccccc}
\tablenum{1}
\tablewidth{0pt}
\tablecaption{Parameters of the Loops (continued)}
\tablehead{
\colhead{Model} & \colhead{$L$} & \colhead{$z_{\rm TR}$} & \colhead{$\Gamma$} & \colhead{$B_{\rm min}$}
&\colhead{$B_{\rm TR}$} &\colhead{$p$} & \colhead{$\tau_{AC}$} &\colhead{$m$} &\colhead{$Q_{\rm TR}$} &
 \colhead{$\eta_{Q}$} \\
\colhead{} & \colhead{[Mm]} & \colhead{[km]} & \colhead{} & \colhead{[G]}
& \colhead{[G]} & \colhead{[$\rm dyne/cm^{2}$]} & \colhead{s} &
\colhead{} & \colhead{[$\rm erg/cm^{3}/s$]}
&\colhead{} } 
\startdata
f18r1 & 99.3 & 3230 & 17.6 & 41.6  & 733.8   & 0.25 &13.29 &0.679   &$3.169\times10^{-3}$ &7.02\\
f18r2 & 100.7& 2624 & 19.9 & 41.6  & 828.8   & 1.00 &22.83 &0.555   &$4.570\times10^{-3}$ &5.26\\
f18r3 & 102.0& 2015 & 22.6 & 41.6  & 941.5   & 4.00 &38.21 &0.723   &$6.225\times10^{-3}$ & 9.53\\
\hline
f19r1 & 174.3 & 3226 & 10.5 & 8.7  & 91.4    & 0.25 &101.66 &0.768  &$8.478\times10^{-4}$ & 6.10\\
f19r2 & 177.6 & 2620 & 11.9 & 8.7  & 103.2   & 1.00 &173.06 &0.732  &$1.035\times10^{-3}$ & 6.13\\
f19r3 & 182.0 & 2014 & 15.3 & 8.7  & 132.5   & 4.00 &287.69 &0.786  &$1.612\times10^{-3}$ & 8.52\\
\hline
f20r1 & 244.6 & 3227 & 27.9 & 3.7  & 102.7   & 0.25 &285.72 &1.014   &$1.285\times10^{-3}$ & 29.23\\
f20r2 & 248.5 & 2631 & 33.8 & 3.7  & 124.5   & 1.00 &484.95 &1.047   &$1.895\times10^{-3}$ & 39.83\\
f20r3 & 252.0 & 2033 & 42.6 & 3.7  & 156.8   & 4.00 &803.13 &1.069   &$2.845\times10^{-3}$ & 55.05\\
\hline
f21r1 & 198.7 & 3223 & 15.8 & 5.9  & 93.1     & 0.25 &163.56 &0.803   &$6.909\times10^{-4}$ & 9.19\\
f21r2 & 201.4 & 2620 & 16.9 & 5.9  & 99.4     & 1.00 &275.60 &0.857   &$9.515\times10^{-4}$ & 11.27\\
f21r3 & 204.5 & 2016 & 18.9 & 5.9  & 111.4    & 4.00&454.15 &0.872    &$1.188\times10^{-3}$ & 13.00\\
\hline
f22r1 & 190.3 & 3226 & 23.1 & 3.8  & 88.5     & 0.25 &210.70 &0.878   &$7.567\times10^{-4}$ & 15.75\\
f22r2 & 192.2 & 2619 & 24.8 & 3.8  & 94.8     & 1.00 &358.21 &0.863   &$9.505\times10^{-4}$ & 15.93\\
f22r3 & 194.1 & 2015 & 26.1 & 3.8  & 100.0    & 4.00 &592.90 &0.889   &$1.187\times10^{-3}$ & 18.18

\enddata               
\end{deluxetable*}

For each model we simulate the 3D dynamics of the Alfv\'{e}n
waves in the neighborhood of the selected field line. 
The waves are described with modest
spatial resolution in the direction perpendicular to the mean magnetic
field ($\ell_\perp \sim 100$ km in the corona). The waves are launched
from the photosphere, and are simulated for a period of 3000~s. The
time resolution is typically about 0.1~s.

\begin{figure*}
\epsscale{1.13}
\plotone{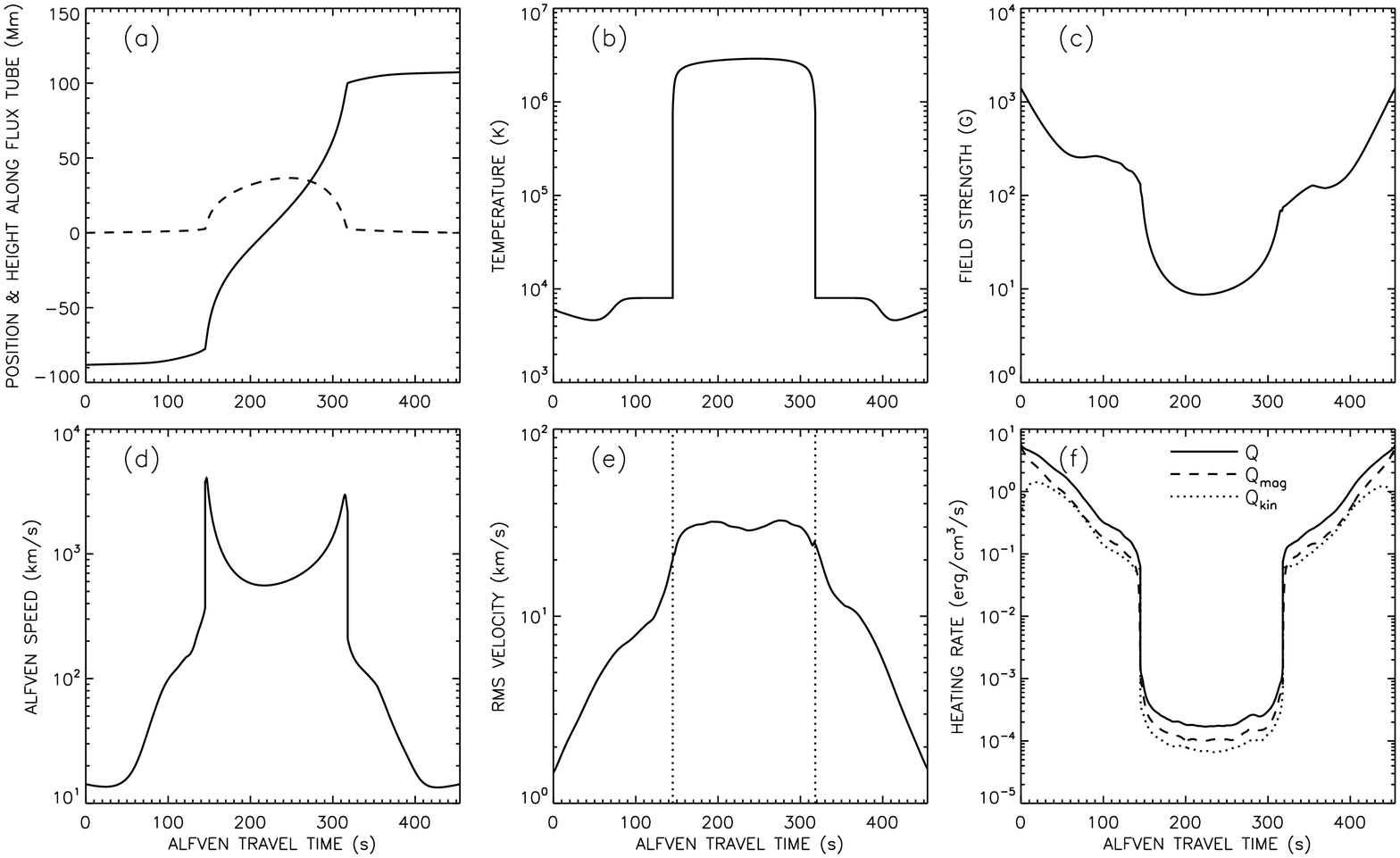}
\caption{Results from numerical simulations of Alfv\'{e}n wave
turbulence in the coronal loop F19 (for model f19r2 in Table 1). Various
quantities are plotted as function of Alfv\'{e}n travel time $\tau$
along the loop: (a) position $s$ (solid curve) and height $z$ (dashed
curve); (b) temperature $T(s)$; (c) background field strength $B(s)$;
(d) Alfv\'{e}n speed $v_A (s)$; (e) wave velocity amplitude
$v_{\rm rms}(s)$; (f) heating rate $Q(s)$.}
\label{fig2}
\end{figure*}

To illustrate the simulation results, Figure~\ref{fig2} shows various
parameters as function of position along the flux tube for model
f19r2 (see Table 1). The model name f19r2 contains the field line number and
the run number refers to different values for coronal pressure. 
Positions are expressed in terms of the Alfv\'{e}n travel time
$\tau$ along the loop:
\begin{equation}
\tau (s) \equiv \int_{s_0}^s \frac{ds^\prime} {v_A (s^\prime)} ,
\label{eq:tau}
\end{equation}
where $s_0$ denotes the position of the positive polarity footpoint.
Figure~\ref{fig2}(a) shows the position $s$ (solid curve) and height
$z$ (dashed curve) plotted as functions of $\tau$. Figure \ref{fig2}(b)
shows the temperature $T(s)$, which varies from about 6000 K in the
photosphere and 8000 K in the chromosphere to about 1 MK in the
corona. The magnetic field strength $B(s)$ varies from 1400 G at the
base of the photosphere to 8.7 G in the corona (see
Figure~\ref{fig2}(c)). The values of $\tau$ on the lower axes of these
plots indicate that the time for an Alfv\'{e}n wave to propagate from
the base of the photosphere to the base of the corona is about 145 s
at the positive polarity end of the loop (left side of each figure)
and 136 s at the negative polarity end (right side). This is
comparable to the time of 173 s for the wave to travel the much larger
distance through the corona ($L = 178$ Mm). Therefore, to understand
Alfv\'{e}n wave dynamics in coronal loops it is important to include
the lower atmospheres at the two ends of a loop. Figure~\ref{fig2}(d)
shows that the Alfv\'{e}n speed varies from about 15 km s$^{-1}$
in the photosphere to more than 1000 km s$^{-1}$ in the low
corona. Therefore, waves traveling to the corona suffer strong wave
reflection \citep[][]{Hollweg1981}. Nevertheless, we find that the waves in the corona build
up to significant amplitudes and have strong nonlinear interactions.

Figure \ref{fig2}(e) shows the time-averaged, root-mean-square velocity
of the waves, $v_{\rm rms} (s)$, as predicted by the reduced MHD
model. The velocity is averaged over the cross-section of the flux
tube and over the period $200 < t < 3000$ s. Note that $v_{\rm rms}$
increases from about 1.5 km s$^{-1}$ in the photosphere (the
imposed footpoint velocity) to about 35 km s$^{-1}$ in the
corona. This amplification is mainly due to the stratification of
density, which drops by six orders of magnitude over these layers.
The great advantage of the reduced MHD model is that it can easily
deal with such large density differences. The time-averaged heating
rate $Q(s)$ is shown in Figure \ref{fig2}(f), together with the
magnetic and kinetic contributions to the heating. Note that the
heating rate in the  coronal part of the loop is about $10^{-3}$
erg cm$^{-3}$ s$^{-1}$,
much smaller than the heating rates in
the lower atmospheres at the two ends of the loop. In this paper we
focus on the heating in the coronal part
of the loop. The coronal heating rate can be fit to a power law:
\begin{equation}
Q(s) = Q_{\rm TR} \left[ \frac{B(s)} {B_{\rm TR}} \right]^m ,
\label{eq:Qsingle}
\end{equation}
where $Q_{\rm TR}$ is the average heating rate at the two TRs. The
exponent $m$ and heating rate $Q_{\rm TR}$ are determined by the fit.

\begin{figure}
\epsscale{1.10}
\plotone{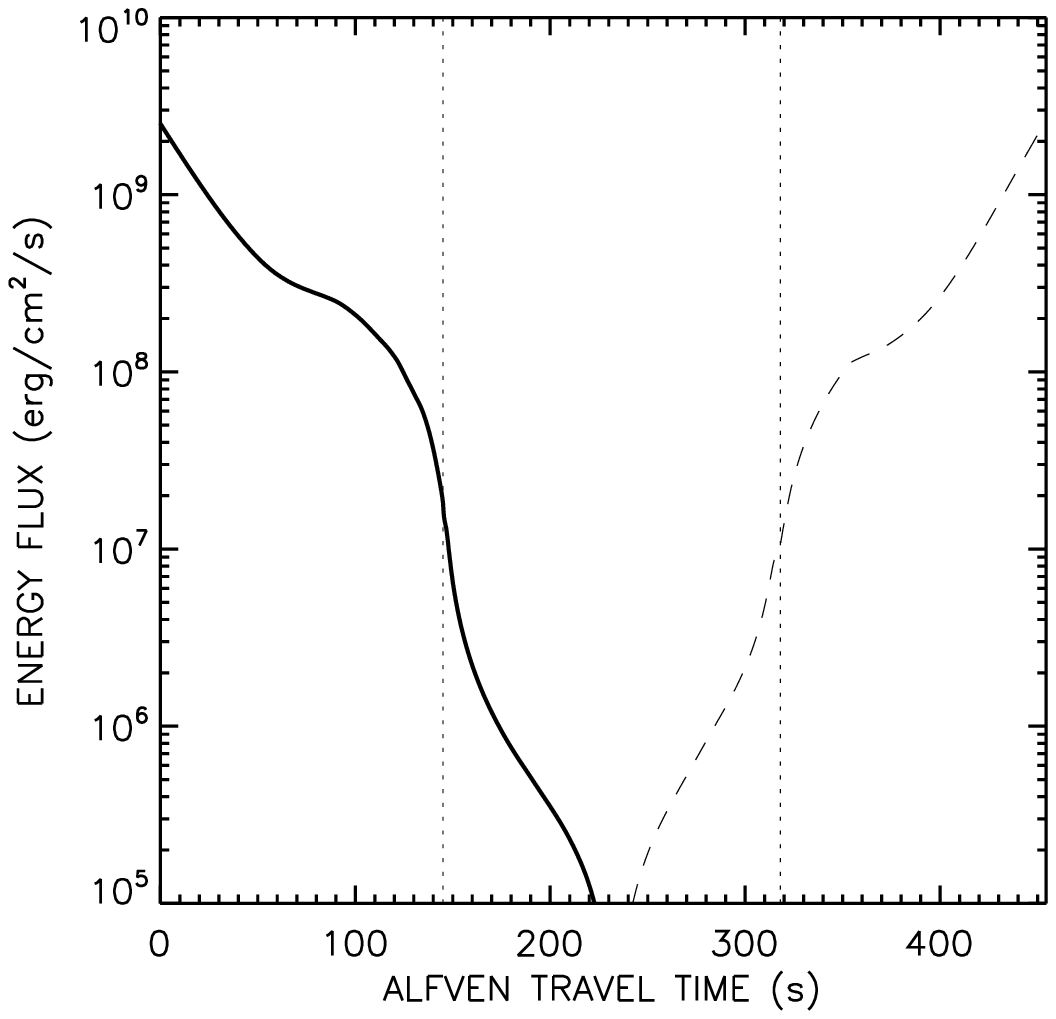}
\caption{The  Alfv\'{e}n wave energy flux $|F_A|$ as function of
  position along the loop for model f19r2. The dotted vertical lines
  indicate the positions of the TRs. Solid and dashed curves indicate
  where $F_A > 0$ and $F_A < 0$, respectively. 
  Note that the energy flux in the
  photosphere $F_A \sim 10^{9}$ $\rm erg~cm^{-2} ~ s^{-1}$ 
  and the flux entering the corona through the TRs is
$F_A \sim 10^{7}$ $\rm erg~cm^{-2} ~ s^{-1}$.}
\label{fig3}
\end{figure}

Figure \ref{fig3} shows the Alfv\'{e}n wave energy flux $F_A (s)$ for
model f19r2, as computed with Equation (C12) of paper I. 
The non-radiative energy flux entering the corona is $F _A \sim 10^{7}$
erg cm$^{-2}$ s$^{-1}$
consistent with \citet[]{Withbroe1977}. This energy enters a
coronal loop at both ends and is dissipated within the loop. 
For most models only 5--7\% of the
available energy is dissipated in the corona; the remainder is
dissipated in the photosphere and chromosphere at the two ends of the
coronal loop.

Results similar to those shown in Figure \ref{fig2} are obtained for
all 77 models. This modeling provides information on the turbulent
heating produced by the waves along the 22 selected field lines for
different values of coronal pressure $p$. The model
parameters are listed in Table 1. The parameters include the coronal
loop length $L$, transition region height $z_{\rm TR}$,  coronal loop 
expansion factor $\Gamma$, minimum field strength $B_{\rm min}$ in 
the corona, average field strength $B_{\rm TR}$ at the two TRs, and 
coronal pressure $p$ used in setting up the background atmosphere,
$\tau_{\rm AC}$ is the coronal Alfv\'{e}n travel time between the two
TRs, $\tau_{\rm AC}=\tau(s_{\rm TR2}) -\tau(s_{\rm TR1})$ 
(see Equation (\ref{eq:tau})), exponent $m$ from Equation
(\ref{eq:Qsingle}), the average heating rate $Q_{\rm TR}$ and the
heating ratio $\eta_{Q}$ (see Equation \ref{eq:etaq1}).
In most cases the exponent $m > 0.5$, so the heating rate $Q(s)$
decreases significantly from the legs of the coronal loop where the
magnetic field is strongest to the top where it is weakest.

\section{Developing a Formula for the Heating Rate}

The purpose of this section is to develop a formula describing the
spatial distribution of the time-averaged heating rate due to
Alfv\'{e}n wave turbulence in the observed active region. The formula
is intended to be used in future studies of both thermally stable and
unstable coronal loops, and therefore must describe not only how the
heating depends on position along each field line, but also how it
varies from one field line to another. The formula will be derived
from the numerical results described in the previous section (see
Table 1), therefore, it is based on a realistic model for the 3D
magnetic structure of the observed active region. The simulations
provide heating rates $Q_n (s)$ for all models, where $n$ is the model
index ($n = 1, \cdots , N$), and $N$ is the total number of models
($N = 77$). The analysis proceeds in two steps. First, we consider the
dependence of the heating rate on position along a loop. Then, we
consider the variations between loops, i.e., the dependence of the
heating on loop parameters such as coronal loop length $L_n$ and
coronal pressure $p_n$.

\subsection{Variations Along Field Lines}

\begin{figure}
\epsscale{1.11}
\plotone{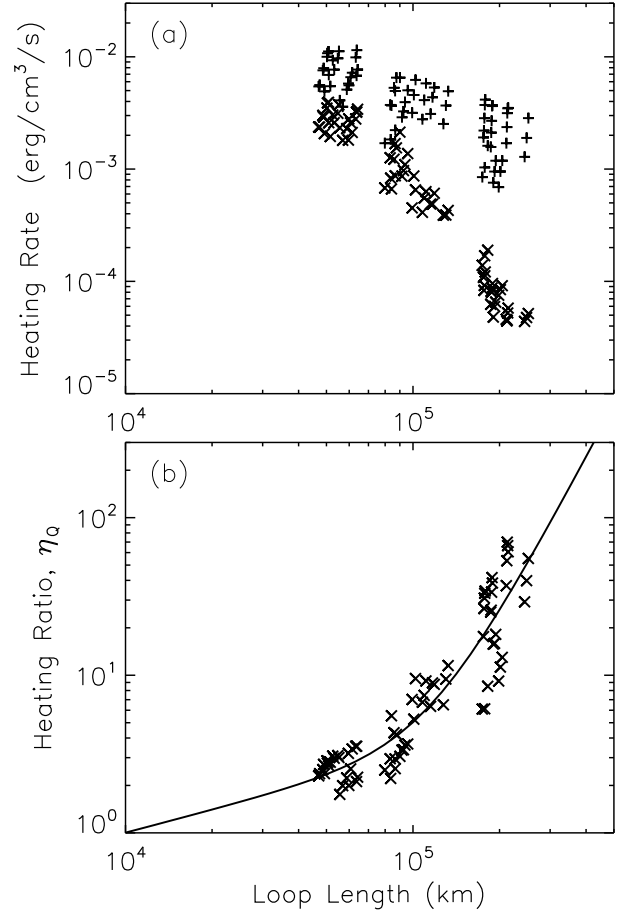}
\caption{(a) Heating rates at the TR, $Q_{{\rm TR}, n}$ (plusses), and
at the loop top $Q_{{\rm min}, n}$ (crosses) are  plotted as a function of the
  loop length $L_n$ for the 77 models. (b) Heating ratio $\eta_{Q,n}$
  for the 77 models plotted as a function of the loop length
  $L_n$.}
\label{fig4}
\end{figure}

Our simulation results indicate that the heating
rate $Q_n (s)$ depends on position along the loop. Figure
\ref{fig4}(a) shows the heating rates at the TR, $Q_{{\rm TR}, n}$ and
at the loop top, $Q_{{\rm min},n}$, plotted as function of loop length
$L_n$ for all 77 models. Note that $Q_{\rm TR}$ is much larger than
$Q_{\rm min}$, and that both quantities decrease with loop length but
$Q_{\rm min}$ drops faster than $Q_{\rm TR}$. The dependence of 
$Q(s)$ on position is usually well described
by a power law in terms of the magnetic field strength $B_n (s)$,
Equation (\ref{eq:Qsingle}). The magnitude of this variation can be
parametrized by
\begin{equation}
\eta_{Q,n} \equiv Q_{{\rm TR},n} / Q_{{\rm min},n} , \label{eq:etaq1}
\end{equation}
where $Q_{{\rm TR},n}$ is computed from a fit as described in section
4, and $Q_{{\rm min},n}$ is the minimum heating rate obtained from the
same fit with $B(s) = B_{min}$. The exponent $m$ in the
relationship between $Q_n(s)$ and $B_n(s)$ is given by 
\begin{equation}
m_n = \frac{\log \eta_{Q,n}} {\log \Gamma_n} . \label{eq:mn}
\end{equation}
where $\Gamma_n \equiv B_{{\rm TR},n} / B_{{\rm min},n}$ is the
expansion factor for model $n$. The ratio $\eta_{Q,n}$ has been
measured for each of the 77 models listed in Table 1. Figure
\ref{fig4}(b) shows this quantity plotted as function of coronal loop
length $L_n$. Note that there is a strong trend with longer loops
having larger $\eta_{Q,n}$ ratios, i.e., stronger variations of
heating rate $Q_n(s)$ along the loop. We also plotted $\eta_{Q,n}$
against other loop parameters, but found that the loop length
$L_n$ provides the strongest correlation.

The simulation results have been fitted with the following formula:
\begin{equation}
\eta_Q (L_n) = a_0 L_{5,n}^{a_1} + a_2 L_{5,n}^{a_3} , \label{eq:etaq2}
\end{equation}
where $L_{5,n}$ is the loop length in units of $10^5$ km for model
$n$. We also assume that short loops (not simulated) must have nearly
uniform heating, so we impose the further constraint that $\eta_Q
\approx 1$ for $L = 10^4$ km and the Equation (\ref{eq:etaq2}) only applies
to loops with $L\ge 10^4$ km. The resulting fit has the following
values for the coefficients:
\begin{equation}
a_0 = 3.021 , ~~~ a_1 = 0.479 , ~~~ a_2 = 2.014, ~~~ a_3 = 3.442 ,
\label{eq:fit}
\end{equation}
and is shown by the solid curve in Figure \ref{fig4}b. Together with
the PFSS model, Equations (\ref{eq:etaq2}) and (\ref{eq:fit}) can be
used to predict the spatial variation of the heating ($\eta_Q$) for
any field line in the observed active region.

\subsection{Variations Across Field Lines}

We now consider all spatial variations of the heating rate $Q$ in
the corona, including variations across magnetic field lines.
Different formulae for $Q$ with different dependencies on the
coronal loop parameters were tested. As an example, let us consider
the case where $Q$ depends on the coronal loop length $L$ and
coronal pressure $p$. In this case the heating rate could be written
in the following logarithmic form: 
\begin{equation}
\log Q(s) = C_0 + C_1 \log L_{5} + C_2 \log p
+ m(L, \Gamma) \log \left[ \frac{B(s)} {B_{\rm TR}} \right] ,
\label{eq:Q}
\end{equation}
where $L_5$ is the coronal loop length (in units of $10^5$ km),
$p$ is the coronal pressure (in $\rm dyne ~ cm^{-2}$), and $m(L, \Gamma)$
is computed in accordance with Equations (\ref{eq:mn}) and
(\ref{eq:etaq2}):
\begin{equation}
m(L, \Gamma) = \frac{\log (a_0 L_{5}^{a_1} + a_2 L_{5}^{a_3})} {\log \Gamma} .
\label{eq:m}
\end{equation}
The last term in Equation (\ref{eq:Q}) describes the variations of
the heating rate {\it along} the field lines, and is consistent with
Equation (\ref{eq:Qsingle}).

\begin{figure}
\epsscale{1.13}
\plotone{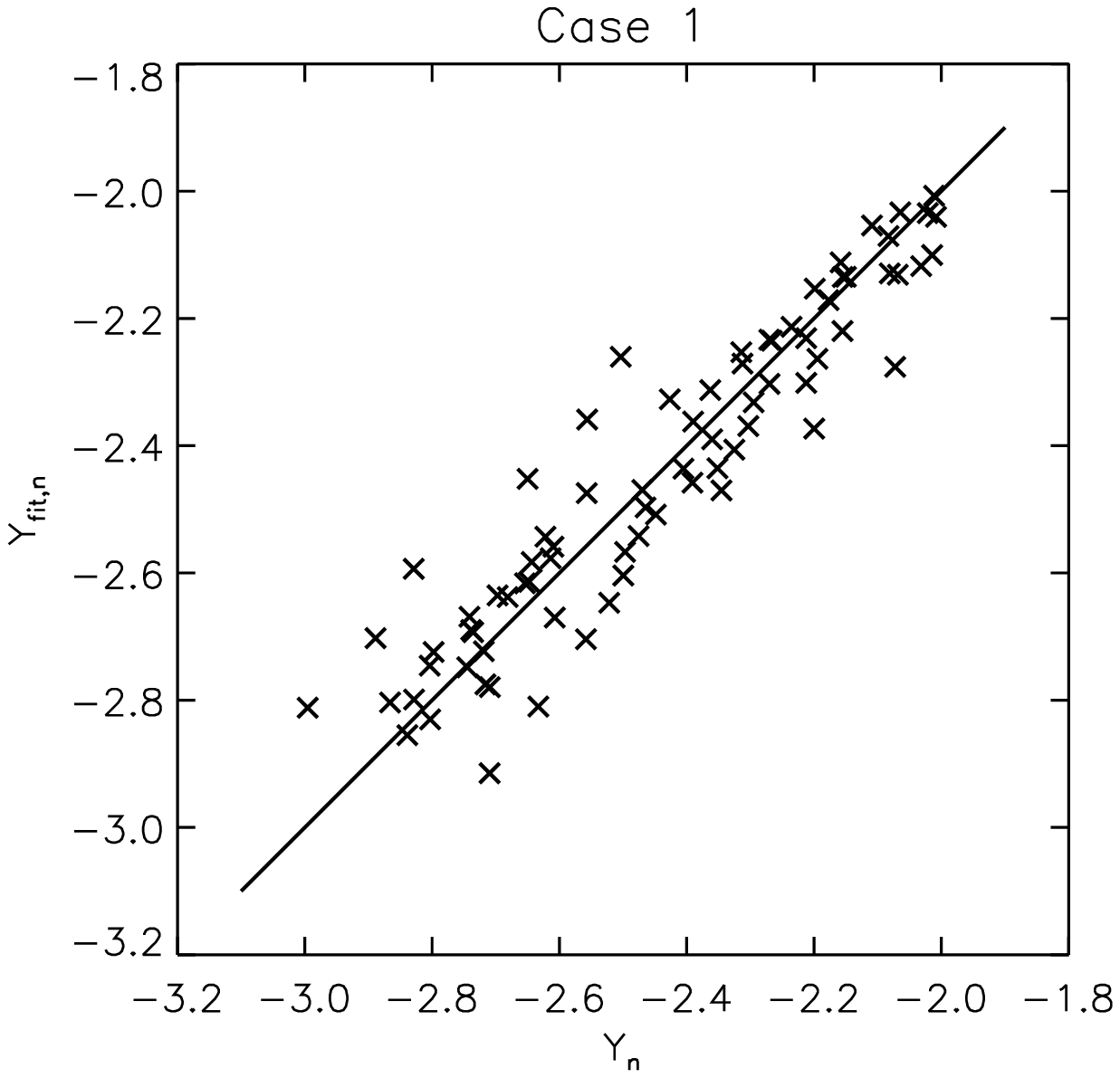}
\caption{The result for $\overline{Y}_{{\rm fit},n}$ is plotted as a
  function of $\overline{Y}_n$.}
\label{fig5}
\end{figure}

We now describe how the constants $C_0$, $C_1$ and $C_2$ in Equation
(\ref{eq:Q}) are determined. Bringing the last term on the left-hand
side of the equation, and applying the formula to model $n$, we can
define 
\begin{equation}
Y_n (s) \equiv \log Q_n (s) - m(L_n, \Gamma_n) \log \left[ \frac{B_n (s)}
{B_{{\rm TR},n}} \right] , \label{eq:Ys}
\end{equation}
where $Q_n(s)$ and $B_n(s)$ are the heating rate and magnetic field
strength as functions of position along the loop, $B_{{\rm TR},n}$ is
the field strength at the TR, and $L_n$ is the loop length. By
construction $Y_n (s)$ is nearly constant along the coronal part of
the loop, therefore, we average it over the coronal grid points to
obtain a single value $\overline{Y}_n$ for each model. Then we fit the
following formula to the measured values of $\overline{Y}_n$:
\begin{equation}
\overline{Y}_{{\rm fit},n} = C_0 + C_1 \log L_{5,n} + C_2 \log p_n ,
\label{eq:Yfit}
\end{equation}
where $L_{5,n}$ is the loop length (in units of $10^5$ km), and
$p_n$ is the coronal pressure (in $\rm dyne ~ cm^{-2}$). The
coefficients $C_0$, $C_1$ and $C_2$ are computed by minimizing
the following quantity:
\begin{equation}
\chi^2 = \frac{1}{N} \sum_{n=1}^{N} \left( \overline{Y}_{n} -
\overline{Y}_{{\rm fit},n} \right)^2 ,
\end{equation}
which is a standard regression analysis. In the present case we find:
\begin{equation}
C_0 = -2.43, ~~~~
C_1 = -0.95, ~~~~
C_2 = 0.19 .
\end{equation}
The error of the fit for a particular model is $\Delta \overline{Y}_n
= \overline{Y}_{{\rm fit},n} - \overline{Y}_{n}$, and averaged over
models the error equals the minimum value of $\chi$. 
The variances of the fit are defined by 
\begin{equation}
\sigma_{1}^2  \equiv  C_{1}^2  \frac{1}{N} \sum_{n=1}^{N} \left( X_{1n} -
\overline{X_1}  \right)^2   = 0.05,
\label{eq:ver1}
\end{equation}
and 
\begin{equation}
\sigma_{2}^2  \equiv  C_{2}^2  \frac{1}{N} \sum_{n=1}^{N} \left( X_{2n} -
\overline{X_2}  \right)^2   = 0.0087,
\label{eq:ver2}
\end{equation}
where $X_{1n}=\log L_{5,n}$,  $X_{2n}=\log p_n$, and $\overline{X_1}$ and $\overline{X_2}$   
are the averages of these quantities. Figure \ref{fig5}
shows the relationship between $\overline{Y}_n$ and 
$\overline{Y}_{{\rm fit},n}$. Note that the average error of the fit
is less then 0.1 in the base-10 logarithm. The scatter in Figure
\ref{fig5} is due to several effects. First, the above Equation 
(\ref{eq:Q}) is only an approximation and does not fit the numerical
results perfectly. Second, there are real differences between loops 
that have the same length and pressure. For example, f9r3 and
f19r2 have the same loop length and coronal pressure, but the heating
for f9r3 is a factor of 2.75 larger than that for f19r2. This is due
to the fact that field line F9 is rooted in strong magnetic field
(sunspot), while F19 is low-lying and rooted in weaker fields.      
 
In the above discussion we used the coronal loop length $L$ and
pressure $p$ as an example, but the same approach can be used with any
combination of loop parameters. 

We examined the dependency of the heating rate on 
different parameters of the field line. The result is illustrated in Table 2. 
The first four rows show the cases with two parameters from 
the selection of $L$, $p$, $\rho_{\rm min}$, and $\tau_{\rm AC}$. 
The last four rows are cases with three model parameters 
including the averaged transition region magnetic field strength $B_{\rm TR}$.
$C_0$ is the constant in the heating equation (Equation (\ref{eq:Q})). 
The exponents in the equation are $C_1$, $C_2$, and $C_3$. The variances 
are ${\sigma_1}^2$, ${\sigma_2}^2$, and ${\sigma_3}^2$.
The fit error $\chi$ is represented in the last column. 

\begin{deluxetable*}{crcccccccccc}
\tablenum{2}
\tablewidth{0pt}
\tablecaption{Testing the parameter dependence of the coronal heating Equation (\ref{eq:Q})}
\tablehead{
\colhead{Case} & \colhead{Par1} & \colhead{Par2} & \colhead{Par3} &
\colhead{$C_{0}$} & \colhead{$C_{1}$}
&\colhead{$C_{2}$} &\colhead{$C_{3}$} & \colhead{$\sigma_{1}^2$} &\colhead{$\sigma_{2}^2$} &\colhead{$\sigma_{3}^2$}&
 \colhead{$\chi$} }
\startdata

1 &  $L$     & $p$  & - & -2.43   & $-0.95 \pm 0.24$ & $0.19\pm 0.12$ &-&0.05  & 0.0087  & - &0.092\\
2 &  $L$    &$\rho_{min}$ & - & -2.28  & $-0.81\pm 0.26 $& $0.24 \pm 0.15$&- & 0.04&0.01&-&0.093\\
3 &  $\tau_{AC}$ & $p$ & - & -2.60   & $-0.38\pm 0.09$  & $0.32 \pm 0.12$   & -  & 0.05  &  0.02 & - & 0.10   \\
4 &  $\tau_{AC}$ & $\rho_{min}$ & - &  -2.31    & $-0.30\pm0.09 $  & $0.39\pm 0.14$ & -  & 0.03  &  0.03 & - & 0.089   \\
\hline
5 & $L$ & $p$ & $B_{TR}$ & -2.44     & $-0.95 \pm 0.25$ &$0.19\pm 0.12$ & $0.01\pm 0.17$ & 0.05  & 0.009 & 0.00002 & 0.091\\
6 & $L$ & $\rho_{min}$ & $B_{TR}$ & -2.30   &$-0.80\pm 0.27$ & $0.24\pm 0.15$ & $0.04\pm 0.17$ & 0.03  & 0.01 & 0.0002 & 0.092\\
7 & $\tau_{AC}$ & $p$ & $B_{TR}$ & -2.51  & $-0.43 \pm 0.11$  & $0.35 \pm 0.12$  & $-0.19 \pm 0.18$& 0.06  & 0.03 & 0.005 & 0.082\\
8 & $\tau_{AC}$ & $\rho_{min}$ & $B_{TR}$ & -2.27  &$-0.31\pm 0.10$ & $0.39\pm 0.14$ & $-0.09\pm 0.18$ & 0.04  & 0.03 & 0.001 & 0.085
\enddata
\end{deluxetable*}
  
From Table 2, case 1 is described in detail in this section. In case
2, we replace the gas pressure $p$ with minimum density
$\rho_{min}$. The exponent $C_1$ doesn't change very much. In fact
$C_1$ is always negative for all cases, pointing to the fact that the
heating rate varies inversely with the length of the loop. The
exponent $C_2$ is positive, confirming the direct correlation between
the heating rate and gas pressure. 
In cases 3 and 4, we substitute the length of the loop with $\tau_{\rm
  AC}$, the exponents $C_1$ and $C_2$ have similar values compared to
the cases 1 and 2. The values of $\chi^2$ indicate that the models 1 to 4 are equally good.

For cases 5 to 8, we include $B_{\rm TR}$. The interesting point is
that the heating structure does not change significantly due to the effect of the
inclusion of $B_{\rm TR}$. The exponent $C_3$ corresponding to $B_{\rm TR}$ is very
small and the error in the calculation of this term is larger than the
exponent itself. This shows that the equation of the heating is better
described by the cases 1 to 4. Apparently, the
magnetic field strength dependency of the heating rate equation is
well described by the term $ B(s)/ {B_{\rm TR}} $, and
including the term $B_{\rm TR}$, doesn't improve the fit.

\section{Heating Rate Variations Within a Coronal Loop}
\label{sect:six}

Papers I and II demonstrated that our model of Alfv\'{e}n wave
turbulence predicts that the heating rate $Q$ varies strongly with
space and time within a given loop. 
Here we explore ``macroscopic'' variations in $s$ and $t$ (averaged
over $x$ and $y$) in section 6.1, and then smaller ``microscopic''
variations resolved over the loop cross section in section 6.2.

\subsection{Variations over Time and Loop Length}
 \label{sect:sixone}

\begin{figure}
\epsscale{1.06}
\plotone{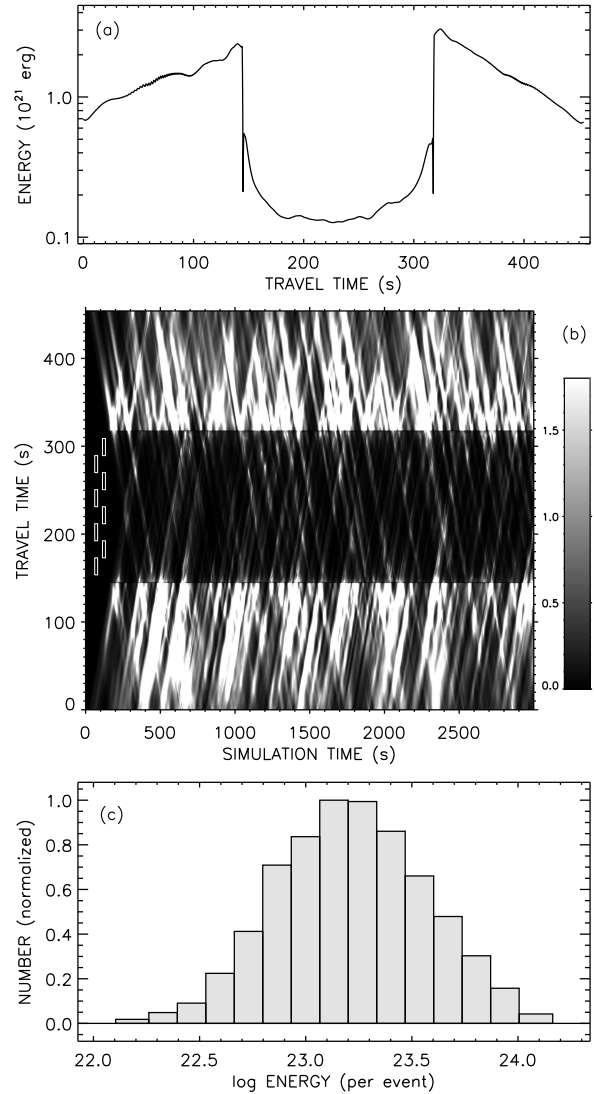}
\caption{Spatial and temporal variations of the heating rate $Q(s,t)$
for coronal loop F19 as a function of position $s$ and time $t$.
(a) Energy as a
function of position along the loop averaged over time.  (b)
Dependence of energy on position (vertical axis) and time
(horizontal axis). The strongest
heating events occur when two ridges intersect, indicating
where counter-propagating wave pulses interact with each other. (c)  
The statistical distribution of energy.}
\label{fig6}
\end{figure}

The Alfv\'{e}n wave turbulence model predicts that the heating rate
$Q$ varies strongly in space and time. We study the spatial 
and temporal intermittency of heating 
events along a representative coronal loop by computing the total 
energy released by heating in each discrete zone of the model. In 
other words, for the same model f19r2 shown in Figure \ref{fig2}, 
we compute the energy per zone 
\begin{equation}
E(s,t) =Q(s,t) \, \Delta s\, \pi R^{2} \, \Delta t
\end{equation} 
where $\Delta s$ is the distance between neighboring loop length 
grid zones (which varies as a function of $s$), $\pi R^{2}$ is the 
cross sectional area of the loop, and $\Delta t = 1.08$~s is the 
interval between output time steps. 
Figure \ref{fig6} (top panel) shows the energy per zone as a function of
position along the flux tube, averaged over time in the same  
way as was done in Figure \ref{fig2}.
Figure \ref{fig6} (middle panel) illustrates the dependence of $E$ 
on position $s$ (plotted as Alfv\'{e}n wave travel time on the
vertical axis) and time $t$ (on horizontal axis).
The gray-scale image shows $E(s,t)$ on a linear scale in units of 
$10^{21}$ erg. 
Because the vertical axis is given in terms of  Alfv\'{e}n 
wave travel time, the trajectories of individual wave ``packets'' 
are straight lines, and it is clear that the strongest heating events 
occur where upward and downward packets collide with one another.

Figure \ref{fig6} (left side of middle panel) also shows representative
``boxes'' that are defined to extend for 19.4 s in simulation time
and 19.4 s in Alfv\'{e}n wave travel time. 
This box size was chosen because it appears to encompass the 
duration and spatial extent of most individual discrete heating 
events that appear in the coronal part of the simulations. 
In order to study the intermittency of nanoflare-like energy release
events, we binned the coronal part of the simulation into a grid 
of boxes of this size.
There were 8 boxes along the loop (the first starting at travel time
153.6 s; the last ending at travel time 308.5 s), and there were 
141 boxes in the time domain (the first starting at $t=248.4$ s; the
last ending at $t=2988.1 s$).
Thus, there were a total of 1128 boxes, each containing a given amount
of energy dissipation. 
The results discussed here do not depend on our choice of starting or
ending locations for the definition of the box edges. 

The distribution of summed energy values per event (i.e., per box)
range over two orders of magnitude, with a minimum value of 
$1.3 \times 10^{22}$ erg, a maximum value of $1.4 \times 10^{24}$ 
erg, and a mean value of $2.2 \times 10^{23}$.
These values are highly reminiscent of what is expected for classical
nanoflares \citep[][]{Hannah2011}. 
Figure \ref{fig6} (bottom panel) shows the statistical distribution of 
energy values collected into bins of equal logarithmic spacing. 
The roughly Gaussian appearance of the distribution suggests a
{\em lognormal} distribution of energy release events along this 
particular loop.
This kind of distribution has been suggested both for statistics 
of solar irradiance variability \citep[][]{Pauluhn2007} and for 
magnetic field variations in the heliosphere \citep[][]{Burlaga2001}.
However, we believe that the low energy cutoff of the distribution in
this case is due to the limited dynamic range of the present
simulations (i.e., a spatial resolution of about 10\% of the tube
radius).
Thus, we anticipate that when all relevant scales are included, the
distribution of energy per event should become broader and possibly
approach a power-law shape as has been found to occur in many other
forms of intermittent energy release. 
  
\subsection{Variations over Loop Cross Section}
 \label{sect:sixtwo}

\begin{figure*}[!t]
\epsscale{1.09}
\plotone{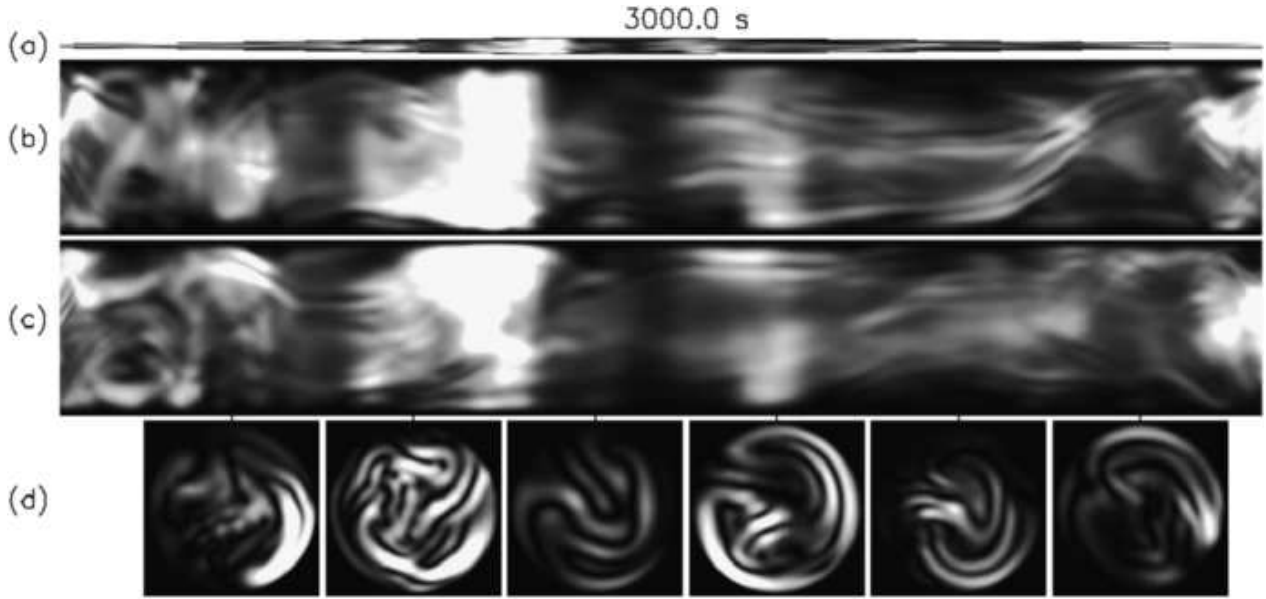}
\caption{The spatial variations of the heating rate in the coronal
  loop F19 (model f19r2). (a) The normalized heating rate integrated
  over the coordinate x. The coronal part of the loop is
  shown. (b) The normalized heating rate integrated over the $x$
  coordinate and is on an expanded scale with
  positions in terms of the Alfv\'{e}n travel time $\tau (s)$. (c) The
  normalized heating rate integrated over the $y$ coordinate. (d) The
  normalized heating rate at six different cross-sections along the loop.}
\label{fig7}
\end{figure*}

To understand how the plasma inside the loop is heated, it is
important to consider the {\it local} heating rate $Q(x,y,s,t)$,
which is also a function of the transverse coordinates $x$ and $y$.
This rate is computed using expressions (\ref{eq:Qkin2}) and 
(\ref{eq:Qmag2}). Figure \ref{fig7} shows results for model f19r2
for a particular instant at the end of the simulation (see also the
animated version included in the on-line supplementary material). 
Figure \ref{fig7}(a) shows the
normalized heating rate integrated over the $x$ coordinate:
\begin{equation}
I(s,y) = \frac{\int Q(x,y,s,t) dx} {R(s) Q(s)} . \label{eq:Isy}
\end{equation}
where $s$ is the horizontal coordinate in the figure, $y$ is the
vertical coordinate in the figure, and $Q(s)$ is the time-averaged
heating rate. In this figure the transverse and longitudinal variations
of the heating are shown on the same spatial scale, but the loop
curvature is neglected. Only the coronal portion of the loop is shown;
the TRs are located at the left and right edges of the diagram.
Figure \ref{fig7}(b) shows the same quantity on an expanded scale
with $s$ replaced by the Alfv\'{e}n travel time $\tau (s)$ and the
$y$ coordinate normalized to the tube radius $R(s)$, and Figure
\ref{fig7}(c) shows a similar plot for the heating integrated over the
$y$ coordinate. In these figures the spatial variations of the heating
within the loop can be seen more clearly. Note that there is a strong
heating pulse at position $s \approx 0.4 L$ from the left TR, and
weaker wave-like disturbances at other places along the loop. 

Figure \ref{fig7}(d) shows the normalized heating rate in six
different cross-sections along the loop, $I(x,y) = Q(x,y,s,t)/Q(s)$.
The positions of these cross-sections are indicated by thin lines
between panels (c) and (d). Note that the heating occurs in thin
ridges and rings that are located near current sheets and shear
layers \citep[][]{Oughton2001, Mininni2009, Servidio2011} , 
i.e., sites of strong magnetic and velocity shear. The shapes
of these rings are different in the different cross-sections, and vary
rapidly with time. In our simulations the ridges cover a significant
fraction of the loop cross-section, so the heating is not highly
localized within the loop. This may be an artifact of our limited
spatial resolution of the waves, together with the fact that we use
hyperdiffusion to describe the wave damping. We have simulated the
Alfv\'{e}n wave turbulence with the use of  hyperdiffusion as in papers
I and II. 

An animated version of Figure \ref{fig7} is included as an on-line 
supplementary material. The animation covers a period of about 432 s near the
end of the 3000-s simulation, and shows the spatio-temporal variations
of the heating $Q(x,y,s,t)$ as seen from the side (panels (a), (b) and
(c)) and in the six cross-sections (panel (d)). The animation shows
wave-like disturbances emerging from the two TRs and propagating into
the corona. The injection of waves into the corona is a highly
intermittent process, which is a consequence of the random nature of
the footpoint motions and the turbulence in the chromosphere. The
strongest brightenings occur when oppositely directed wave packets
``collide'' with each other. This is a result of nonlinear wave--wave
interactions, which produce a rapid cascade to higher wavenumbers and
dissipation of the wave energy. 

\section{Discussion and Conclusions}
\label{sect:Discussion}

In our modeling, we simulated the dissipation and propagation of
Alfv\'{e}n waves in coronal loops. In order to fully understand
Alfv\'{e}n wave dynamics in coronal loops 
we included the lower atmospheres at the two ends of a loop as in
papers I and II. The waves are launched from the photosphere, and 
are simulated for a period of 3000~s.
Along the field lines, the Alfv\'{e}n speed varies from about 10--15  
$\rm km ~ s^{-1}$ in the photosphere to more than 1000 $\rm km ~
s^{-1}$ in the low corona. Therefore, waves traveling to the corona
encounter strong wave reflection. We find that the waves 
in the corona build up to significant amplitudes and have strong
nonlinear interactions. This results in the dissipation of Alfv\'{e}n waves.
As shown in paper II, the dissipation rate of Alfv\'{e}n wave
turbulence is sufficient to produce the observed rates of
chromospheric and coronal heating in active regions.

In this work as well as identifying the mechanism responsible for the energy 
input in the corona and the subsequent heating of the coronal loops, 
we predicted the heating rate based on a photospheric magnetogram and 
derived an equation describing the dependence of the heating rate on
loop parameters. Specifically, we simulated the dynamics of 
Alfv\'{e}n waves for 22 field lines in an active region observed on
2012 March 7.
We looked at the variations of the 
wave heating rate for these 22 field lines with field strength, loop length 
and coronal density. The results for different loops were combined into a 
formula describing the average heating rate $Q(s)$ as function of
position $s$ along the loop. As we show in the Alfv\'{e}n wave modeling of 
the coronal heating, the wave dissipation varies strongly with position along the 
loop. In the heating Equation  (\ref{eq:Q}), the dependence on the
position is represented by the term $B(s)$ where magnetic field
strength is a function of position along the loop and the realistic
models of coronal loops all point to this variation.
As the Equation (\ref{eq:Q}) indicates the heating also depends on the loop parameters such 
as coronal loop length $L$ and coronal pressure $p$.
Results for other combinations of parameters are shown in Table 2. 
The spatial variation of the heating rate is not arbitrarily 
prescribed but is a natural consequence of the Alfv\'{e}n wave
turbulence model. It is important to note that the loops heating
profile plays an important role in determining whether the loops are thermally stable or 
not. 

Equation (\ref{eq:Q}) and similar expressions listed in Table 2 are
intended to be used for constructing 3D time-dependent models of 
coronal plasma in an observed active region, and could also be used for other regions.
Such modeling requires solving the energy transport equation for
coronal loops, i.e., the balance between non-thermal heating,
radiative losses and thermal conduction. In general both thermally
stable and unstable coronal loops will be present, and for the latter
the plasma temperature and density will vary strongly with time. 
Therefore, in the present modeling the coronal pressure $p$ was
treated as a free parameter, and the loops were not constrained to be
in thermal equilibrium. Computations were done for a wide range of 
coronal pressures, so that the heating-rate formulae would accurately
describe the effects of the coronal density on the propagation and 
dissipation of the Alfv\'{e}n waves for a wide range of coronal conditions.
 
\citet[]{Mandrini} investigate the coronal
heating dependence on the magnetic field strength and coronal loop
length. \citet[]{Schrijver2004} compare various models of the heating 
with TRACE observations of the corona, and found that  the heating
flux density $F_{H}$ varies as $B/L$, in agreement with one of the
 models studied by \citet[]{Mandrini} and \citet[]{Demoulin2003}. 
This corresponds to a heating rate per unite volume that depends
strongly on loop length: $Q \propto L^{-2}$.
However, the above mentioned authors do not include the variation
along the loop. In our studies, the
heating rate formula has a weaker dependency on the loop length than
that found by \citet[]{Schrijver2004} and the
magnetic field dependency of our heating expression is well represented by the
term $ B(s)/ {B_{\rm TR}} $ in equation
(\ref{eq:Q}). 

In our present work we do not address the question how the coronal plasma
responds to the heating events. A comprehensive overview of the
coronal heating problem is presented by \citet[]{Klimchuk2006}, who argues
that loops are heated by ``storms'' of nanoflares. \citet[]{Warren2002}
explored the possibility that active region loops are a collection of
small-scale filaments that have been impulsively heated. 
They found that the density variations lag the temperature variations
by several hundred seconds, and during the cooling phase the filaments 
are significantly denser than steadily heated loops with the same
temperature. This is important for understanding the brightness of
observed loops at EUV wavelengths. \citet[]{Reale2005} were the first
to have simulated the response of a loop to nanoflare heating associated 
with MHD turbulence, and they simulated the emission in EUV
passbands. The modeled coronal temperatures and pressures are somewhat
lower than those observed in active regions; we suggest this may be
due to the fact that this model does not include wave amplification in 
the lower atmosphere. Future modeling of Alfv\'{e}n wave turbulence should
take into account the dynamic response of the plasma to the heating
events, including the effects of spatial variations of temperature and 
density over the loop cross-section. 

In the present model the TR is treated as a discontinuity, and energy
exchange between the chromosphere and corona is not included. 
However, thermal conduction and mass flows along the field lines 
are likely to play an important role in the structure of the TR. 
The wave turbulence model predicts that the wave heating rate in the
chromosphere is orders of magnitude larger than that in the corona,
e.g., see Figure \ref{fig2}(f). In reality some of this energy may be injected
into the low corona by spicules or similar dynamic
phenomena \citep[][]{DePontieu2007b, McIntosh2011}. This may
alter the way Alfv\'{e}n waves reflect at the TR, and may increase the
fraction of wave energy dissipated in the corona.

In our future work we are intending to construct models of the thermal
structure of coronal loops heated by Alfv\'{e}n wave turbulence where
we include the effect of conduction and 
radiative losses, and also make use of Doppler shift measurements as an
important constraints on our coronal loop modeling.

\acknowledgements
We thank James Klimchuk and Peter Cargill for discussions related to
this work. We are most grateful to Alex Voss from the School of Computer
Science
at the University of St.\  Andrews for his support with the
computational work, which was funded by the UK's Engineering and
Physical Sciences Research Council (EP/I034327/1). The HMI and AIA data
have been used courtesy of NASA/SDO and the AIA, EVE, and HMI science
teams. 
This project is supported under contract NNM07AB07C from NASA to the 
Smithsonian Astrophysical Observatory (SAO) and SP02H1701R from LMSAL
to SAO.


\begin{thebibliography}{}

\bibitem[Antiochos \& Klimchuk(1991)]{Antiochos1991}  Antiochos, S. K., \& Klimchuk, J. A. 1991, \apj, 378, 372

\bibitem[Antolin \& Shibata(2010)]{Antolin2010} Antolin, P., \& Shibata, K. 2010, \apj, 712, 494

\bibitem[Aschwanden(2005)]{Aschwanden2005}
Aschwanden, M. J. 2005, The Physics of the Solar Corona
(Berlin: Springer), Chapter 9

\bibitem[Asgari-Targhi \& van Ballegooijen(2012)]{Asgari2012}
Asgari-Targhi, M., \& van Balleggoijen, A. A. 2012, \apj, 746, 81

\bibitem[Bhattacharjee \& Ng(2001)]{Bhattacharjee2001}
Bhattacharjee, A., \& Ng, C. S. 2001, \apj, 548, 318

\bibitem[Burlaga(2001)]{Burlaga2001}
Burlaga, L. F. 2001, JGR, 106, 15917

\bibitem[Cho, Lazarian \& Vishniac(2002)]{Cho2002}
Cho, J., Lazarian, A., \& Vishniac, E. T. 2002, \apj, 564, 291

\bibitem[Chitta et al.(2012)]{Chitta2012}
Chitta, L. P., van Ballegooijen, A. A., Rouppe van der Voort, L.,
et al.  2012, \apj, 752, 48

\bibitem[Cranmer et al.(2007)]{Cranmer2007}
Cranmer, S. R., van Ballegooijen, A. A. \& Edgar, R. J. 2007, \apjs,
171, 520

\bibitem[Cranmer(2010)]{Cranmer2010}
Cranmer, S. R. 2010, \apj, 710, 676

\bibitem[De Pontieu et al.(2007a)]{DePontieu2007a} 
De Pontieu, B., McIntosh, S. W., \& Carlsson, M. 2007a, Science, 318, 1574

\bibitem[De Pontieu et al.(2007b)]{DePontieu2007b} De Pontieu, B., Hansteen, V. H., Rouppe van der Voort, L., van Noort, M., \& Carlsson, M. 2007b, \apj, 655, 624

\bibitem[Dmitruk \& Gomez(1997)]{Dmitruk1997}
Dmitruk, P., \& Gomez, D. O. 1997, \apj, 484, L83

\bibitem[ D\'{e}moulin et al.(2003)]{Demoulin2003} 
 D\'{e}moulin, P., van Driel-Gesztelyi, L., Mandrini, C. H., Klimchuk,
 J. A. \& Harra, L. 2003, \apj, 586,592

\bibitem[Fujimura \& Tsuneta(2009)]{Fujimura2009} Fujimura, D. \& Tsuneta, S. 2009, \apj, 702, 1443

\bibitem[Goldreich \& Sridhar(1995)]{Goldreich1995}
Goldreich, P., \& Sridhar, S. 1995, \apj, 438, 763

\bibitem[Goldreich \& Sridhar(1997)]{Goldreich1997}
Goldreich, P., \& Sridhar, S. 1997, \apj, 485, 680

\bibitem[Hannah et al.(2011)]{Hannah2011}
Hannah, I. G., Hudson, H. S., Battaglia, M., Christe, S. Kasparov\'{a}, J.
Krucker, S., Kundu, M. R., \& Veronig, A. 2011, SSRv, 159, 263

\bibitem[Heyvaerts \& Priest(1983)]{Heyvaerts1983}
Heyvaerts, J., \& Priest, E. R. 1983, \aap, 117, 220

\bibitem[Higdon(1984)]{Higdon1984}
Higdon, J. C. 1984, \apj, 285, 109

\bibitem[Hollweg(1981)]{Hollweg1981} Hollweg, J. V. 1981, \solphys, 70, 25

\bibitem[Hollweg(1986)]{Hollweg1986} Hollweg, J. V. 1986, \jgr, 91, 4111

\bibitem[Karpen et al.(2006)]{Karpen2006}  Karpen, J. T., Antiochos, S. K., \& Klimchuk, J. A. 2006, \apj, 637, 531

\bibitem[Klimchuk(2006)]{Klimchuk2006} Klimchuk, J. A. 2006, \solphys, 234, 41

\bibitem[Klimchuk et al.(2010)]{Klimchuk2010} Klimchuk, J. A., Karpen, J. T., \& Antiochos, S. K. 2010, \apj, 714, 1239

\bibitem[Kraichnan(1965)]{Kraichnan1965} Kraichnan, 1965, Phys.\  Fluids,
  8, 1385  

\bibitem[Kudoh \& Shibata(1999)]{Kudoh1999} Kudoh, T., \& Shibata, K. 1999, \apj, 514, 493

\bibitem[Mandrini et al.(2000)]{Mandrini}  Mandrini, C. H., D\'{e}moulin,
  P.,  \& Klimchuk, J. A. 2000, \apj, 530, 999

\bibitem[Matsumoto \& Shibata(2010)]{Matsumoto2010}
Matsumoto, T., \& Shibata, K. 2010, \apj, 710, 1857

\bibitem[McIntosh et al.(2011)]{McIntosh2011}  McIntosh, S. W., De Pontieu, B., Carlsson, M., Hansteen, V., Boerner, P., \& Goossens, M. 2011, \nat, 475, 477

\bibitem[Mininni \& Pouquet(2009)]{Mininni2009}
Mininni, P. D., \& Pouquet, A. 2009, Phys. Rev. E, 80, 025401

\bibitem[Mok et al.(2008)]{Mok2008} Mok, Y., Miki\'{c}, Z., Lionello, R., \& Linker, J. A. 2008, \apj, 679, L161

\bibitem[Moriyasu et al.(2004)]{Moriyasu2004} Moriyasu, S., Kudoh, T., Yokoyama, T., \& Shibata, K. 2004, \apjl, 601, L107

\bibitem[M\"{u}ller et al.(2003)]{Muller2003} M\"{u}ller, D. A. N., Hansteen, V. H., \& Peter, H. 2003, \aap, 411, 605

\bibitem[Oughton \& Matthaeus(1995)]{Oughton1995}
Oughton, S., \& Matthaeus, W. H. 1995, JGR, 100, 14783

\bibitem[Oughton et al.(2001)]{Oughton2001}
Oughton, S., Matthaeus, W. H., \& Dmitruk, P. 2001, \apj, 551, 565

\bibitem[Oughton, Dmitruk \& Matthaeus(2004)]{Oughton2004}
Oughton, S., Dmitruk, P., \& Matthaeus, W. H. 2004, J. Plasma Phys.,
11, 2214

\bibitem[Pauluhn \& Solanki(2007)]{Pauluhn2007}
Pauluhn, A., \& Solanki, S. K. 2007, \aap, 462, 311

\bibitem[Rappazzo et al.(2008)]{Rappazzo2008}
Rappazzo, A. F., Velli, M., Einaudi, G., \& Dahlburg, R. B. 2008, 
\apj, 677, 1348

\bibitem[Reale et al.(2005)]{Reale2005}
Reale, F., Nigro, G., Malara, F. Peres, G. \& Veltri, P. 2005, \apj, 633, 489

\bibitem[Rosner et al.(1978)]{Rosner1978}
Rosner, R., Tucker, W. H., \& Vaiana, G. 1978, \apj, 220, 643

\bibitem[Schrijver et al.(2004)]{Schrijver2004}
Schrijver, C. J., Sandman, A. W., Aschwanden, M. J., \& DeRosa, M. L.
2004, \apj, 615, 512

\bibitem[Servidio et al.(2011)]{Servidio2011}
Servidio, S., Greco, A., Matthaeus, W. H., Osman, K. T. \& Dmitruk,
P. 2011, JGR, 116, A09102

\bibitem[Shebalin et al.(1983)]{Shebalin1983}
Shebalin, J. V., Matthaeus, W. H., \& Montgomery, D. 1983, J. Plasma
Phys., 29, 525

\bibitem[Stenflo(1973)]{Stenflo1973} Stenflo, J. O. 1973, Solar Phys., 32, 41  

\bibitem[Strauss(1976)]{Strauss1976} Strauss, H. R. 1976,
Phys.\  Fluids, 19, 134

\bibitem[Su et al.(2009a)]{Su2009a} Su, Y. N.,
van Ballegooijen, A. A., Lites, B. W., et al. 2009a, \apj, 691, 105 

\bibitem[Su et al.(2011)]{Su2011} Su, Y., Surges, V., van Ballegooijen, A., DeLuca, E., \& Golub, L. 2011, \apj, 734, 53

\bibitem[Testa et al.(2005)]{Testa2005} Testa, P., Peres, G., \& Reale, F. 2005, \apj, 622, 695

\bibitem[Tomczyk \& McIntosh(2009)]{Tomczyk2009} Tomczyk, S., \& McIntosh, S. W. 2009, \apj, 696, 1384

\bibitem[Ulrich(1996)]{Ulrich1996} Ulrich, R. K. 1996, \apj, 465, 436

\bibitem[van Ballegooijen et al.(2011)]{vanB2011} van Ballegooijen, A. A., Asgari-Targhi, M. , Cranmer, S. R., \& DeLuca, E. E. 2011, \apj, 736, 3

\bibitem[Warren et al.(2002)]{Warren2002}
Warren, H. P., Winebarger, A. R. \& Hamilton, P. S.
2002, \apj, 579, L41

\bibitem[Withbroe \& Noyes(1977)]{Withbroe1977}
Withbroe, G. L., \& Noyes, R. W. 1977, \araa, 15, 363

\end{thebibliography}
\end{document}